\documentclass{ws-ijmpd}
\usepackage{subfigure}
\usepackage{cite}
\usepackage{hyperref}
\usepackage{multirow}

\newcommand{\ba}{\begin{eqnarray}}
\newcommand{\ea}{\end{eqnarray}}
\newcommand{\be}{\begin{equation}}
\newcommand{\ee}{\end{equation}}
\newcommand{\bd}{\begin{displaymath}}
\newcommand{\ed}{\end{displaymath}}
\newcommand{\een}{\nonumber\end{equation}}
\newcommand{\bea}{\begin{eqnarray}}
\newcommand{\eean}{\nonumber\end{eqnarray}}
\newcommand{\eea}{\end{eqnarray}}

\def\la{\langle}
\def\ra{\rangle}

\def\eq#1{Eq.~(\ref{#1})}
\def\eqs#1{Eqs.~(\ref{#1})}
\def\fig#1{Fig.~\ref{#1}}

\newcommand{\plotangle}{0}

\def\mcC{{\mathcal C}}
\def\mcJ{{\mathcal J}}

\newcommand{\chipt}{\chi\rm{PT}}

\newcommand{\mev}{\mathrm{MeV}}

\newcommand{\fm}{\mathrm{fm}}
\newcommand{\gap}{\hspace{10pt}}

\begin{document}

\markboth{Karl Jansen}
{DARK MATTER SEARCH AND THE SCALAR QUARK CONTENTS OF THE NUCLEON}

\catchline{}{}{}{}{}
%
%%%%%%%%%%%%%%%%%%%%%%%%%%%%%%%%%%%%%%%%%%%%%%%%%%%%%%%%%%%%%%%%%%%%

\title{
\vspace*{-1.5cm}
\begin{flushright}
DESY 11-154
\end{flushright}
\vspace*{1.5cm}
Dark matter Search and the Scalar Quark Contents of the Nucleon}

\author{Simon Dinter, Vincent Drach, Karl Jansen}

\address{NIC, DESY, Platanenallee 6, 15738 Zeuthen, Germany}

\maketitle

%\begin{history}
%\received{Day Month Year}
%\revised{Day Month Year}
%\comby{Managing Editor}
%\end{history}

\begin{abstract}
We present lattice QCD simulation results from the European 
Twisted Mass Collaboration (ETMC) for the light, strange and charm quark 
contents of the nucleon. These quantities are important ingredients 
to estimate the 
cross-section for the detection of WIMPs as Dark Matter 
candidates. By employing a particular lattice QCD formulation, 
i.e. twisted mass fermions, accurate results of the light and strange scalar contents
of the nucleon can be obtained. In addition, we provide a bound for the 
charm quark content of the nucleon. 
\end{abstract}


\section{Introduction}

Dark Matter is still 
quite mysterious and its nature not clarified. 
Quite popular candidates are weakly interacting massive particles (WIMPs) 
which appear in many models~\cite{Bertone:2004pz}.
There are a number of experiments that attempt to detect
such WIMPs. The underlying mechanism for the WIMP detection 
is that --due to their assumed large mass-- 
they produce a Higgs boson which in turn can interact
with nucleons.
What the experiments are --in principle-- able to detect then is the 
recoil energy of the nucleon struck in this way by a WIMP which 
is flying through the detector. 

Since the Higgs boson is a scalar, the proposed interaction of 
the WIMP and the nucleon is going to occur via the coupling 
of the Higgs boson to the scalar quark content of the 
nucleon.  
The spin independent (SI) elastic cross section for such a process 
reads \cite{Ellis:2008hf}
\be
\sigma_{\rm SI, \chi N} \sim  
\Big\lvert\sum_{q_f} G_{q_f}(m_\chi^2 ) \la N | \bar{q}_f{q_f} |N \ra \Big\rvert^2 
\label{eq:crosssection} 
\ee
where $G_{q_f}$ is the effective coupling constant between a quark of 
flavour $f$ and the WIMP of mass $m_\chi$ and $\la N | \bar{q}_fq_f |N \ra$ denotes 
the scalar content of the nucleon belonging to a quark flavour $f$.
The strength of the coupling $G_{q_f}$ will depend on the particular 
model it is computed in.

It is important to note that changes of the scalar content of the nucleon 
by about 10\% can lead to large changes in the 
cross-section \cite{Ellis:2008hf,Giedt:2009mr} since 
$\la N | \bar{q}_f{q_f} |N \ra$ appears quadratically in eq.~(\ref{eq:crosssection}).
One problem with calculating the cross-section is that 
the scalar quark contents of the nucleon are only poorly determined. 
If chiral perturbation theory ($\chipt$) is used for their 
calculations
one finds for the ratio $y_N$ of the strange to the light scalar contents
\be\label{eq:y_para}
y_N \equiv  \frac{ 2\la N| \bar{s} s |N \ra }{ \la N| \bar{u} u+ \bar{d} d |N \ra}
\ee
a value of $y_N=0.44(13)$\cite{Young:2009ps}.
This suggests that actually the strange quark content of the 
nucleon is surprisingly large. On the other hand, 
from this $\chipt$ evaluation of $y_N$ only an accuracy of about 30\% 
is achieved leading 
to a large uncertainty in the cross-section itself. 
Thus, is seems highly desirable to compute $y_N$ accurately and 
test whether indeed the scalar quark content of the nucleon is as
large as indicated by $\chipt$. 

In principle, lattice QCD provides a tool to reach a precise quantitative 
result for $y_N$. However, also in this framework the computation 
of $y_N$ turns out to be very difficult. One way to calculate $y_N$ is via  
the Feynman-Hellman theorem~\cite{Feynman:1939zz} which uses the quark mass
dependence of the nucleon mass, 
\be\label{eq:sigma_terms_FH}
\sigma_{\pi N} = m \la N|  \bar{u} u+  \bar{d} d |N \ra = m \frac{\partial m_N}{\partial m}  \gap\textmd{and}\gap \sigma_{ssN} = m_s\la N| \bar{s} s |N \ra = m_s \frac{\partial m_N}{\partial m_s},
\ee
where we denote with $m$ the mass-degenerate light up and down quark masses 
and with $m_s$ the strange quark mass. 
However, to determine $y_N$ taking the numerical derivative
requires many simulations at 
different values of the quark masses. 
In addition, it is unclear how sensitive 
the nucleon mass will be on the variation of the quark mass. 

A second approach is to compute the scalar quark contents 
$\la N | \bar{q}_f{q_f} |N \ra$ directly. The disadvantage here is that 
dis-connected (singlet) contributions are to be computed which are 
generically very noisy in lattice calculations and need a very large 
statistics. 

In this proceedings contribution we want to report about a lattice determination 
of $y_N$ which uses a particular formulation of lattice QCD, namely 
the so-called twisted mass fermions taken at maximal twist. 
What is important for the discussion here is first of all that physical 
quantities computed within this approach scale to the continuum limit, 
i.e. sending the lattice spacing $a$ to zero, 
with a rate that is of $O(a^2)$ leading to a rapid approach towards the continuum 
limit. 
Secondly, twisted mass fermions allow for special techniques to compute 
dis-connected diagrams which give a much better signal to noise ratio 
than most other lattice formulations of QCD. In this report we will make 
heavily use of this special property of twisted mass fermions. 
A third advantage of twisted mass fermions (at maximal twist) is that 
they allow for a straightforward
renormalization of $\la N | \bar{q}_f{q_f} |N \ra$ \cite{dinter} 
avoiding any mixing 
between the bare light, strange and charm quark matrix elements.
Although also chiral invariant lattice fermions share the same property, 
these latter kind of lattice fermions are much more computer time
demanding such that twisted mass fermions, from this point of view,  
are clearly advantageous. 

We want to mention that 
there exist already a number of computations of the strange quark 
content of the nucleon, see 
refs.~\cite{Ohki:2009mt,Ohki::2010cw,Collins:2010gr,Bali:2009dz,Freeman:2009pu,Durr:2010ni,Michael:2001bv,Young:2009zb}. 
For a review we refer to ref.~\cite{Young:2009ps}.
However, because of the aforementioned reasons, these calculations 
are affected with rather large statistical errors. It is our goal 
to improve significantly on this situation and provide a more precise 
value of the scalar content of the nucleon for the light and strange 
quarks. In addition, we will give a bound on the charm quark content. 

%
%\begin{figure}[htb]
%\begin{center}
%\hspace{-15pt}\includegraphics[width=0.7\textwidth,angle=\plotangle]{./Figures/Wimp_N_scattering}
%\end{center}
%\caption{Illustration of the contribution to a WIMP-Nucleon low energy scattering process   
%via a Higgs-boson exchange.}\label{fig:wimp_N_scatt}
%\end{figure}

\section{The calculation and the results}

As said above, we use a particular formulation of lattice QCD called 
twisted mass fermions as proposed by Frezzotti and Rossi, 
see refs.~\cite{Frezzotti:2000nk,Frezzotti:2003ni,Frezzotti:2003xj,Frezzotti:2004wz}. 
The European Twisted Mass Collaboration (ETMC) 
--in which this project is embedded--
is using these kind of lattice fermions for a number of years now
and has obtained many physical results already, see e.g. 
refs.~\cite{Alexandrou:2011nr,Feng:2009ij,Feng:2010es,Feng:2011zk,Baron:2009wt,Baron:2010bv} 
and references therein.

In the latest set of simulations, ETMC has incorporated the complete 
first two quark generations in the simulations, meaning that besides 
the light up and down quarks also the strange and charm quarks are taken 
into account as dynamical degrees of freedom \cite{Chiarappa:2006ae,Baron:2010bv,Baron:2010th}. 
This is presently 
worldwide a rather unique setup for lattice simulations and particularly 
important for the problem 
addressed here since we want to evaluate the strange and 
the charm quark contents of the nucleon.  
In order to compute nucleon scalar matrix elements a 
nucleon 2-point function is needed which is given by 
\be\label{eq:C2pts}
C^{\pm }_{N,\rm 2pts}(t -  t_{\rm src}) =   \sum_{\vec{x}}  
{\rm tr} \Gamma^{\pm}  \la \mcJ_{N}(x) \overline{\mcJ_{N}}(x_{\rm src})  \ra ,
\ee
where the subscript $N$ refers to the proton or to the neutron states for 
which the interpolating fields are given by :
\bd
\mcJ^p = \epsilon^{abc} \left( u^{a,T} \mcC\gamma_5 d^b \right) u^c \gap\textmd{and}\gap   \mcJ^n = \epsilon^{abc} \left( d^{a,T} \mcC\gamma_5 u^b \right) d^c.
\ed
The projectors used are $\Gamma^{\pm} = \frac{1\pm\gamma_0}{2}$, and $\mcC$ is the 
charge conjugation matrix. Using discrete symmetries and anti-periodic boundary 
conditions in the time direction for the quark fields, we have  
$C^{+}_{N,\rm 2pts}(t)  = -C^{-}_{N,\rm 2pts}(T-t)$. 

The other input for the computation of the scalar quark content of the 
nucleon is the zero momentum 3-point function  
\be\label{eq:C3pts}
C^{\pm,O_q}_{N,\rm 3pts}(t_{s},\Delta t_{\rm{op}}) =     
\sum_{\vec{x} ,\vec{x}_{\rm op}}   
{\rm tr} \Gamma^{\pm}\la \mcJ_{N}(x) O_q(x_{\rm op}) \overline{\mcJ_{N}}(x_{\rm src})  \ra , 
\ee
where  $O_q$ is an operator having scalar quantum numbers,  
$\Delta t_{\rm op} =t_{\rm{op}} - t_{\rm{src}} $ is the time of insertion 
of the operator, and $t_s=t-t_{\rm src}$ gives the so-called source-sink 
separation. Since we will consider an operator with a non vanishing vacuum 
expectation value, we also define
\be\label{eq:C3pts_vev_sub}
C^{\pm,O_q,\rm{vev}}_{N,\rm 3pts}(t_{s},\Delta t_{\rm{op}}) =           
C^{\pm,O_q}_{N,\rm 3pts}(t_{s},\Delta t_{\rm{op}}) -      
C^{\pm }_{N,\rm 2pts}(t, x_{\rm src})   \sum_{\vec{x}_{\rm op}} \la O_q(x_{\rm op}) \ra\; .
\ee
The scalar quark operators $O_q$ are given by
\be\label{eq:ops_scalar}
O_l = \bar{u}u + \bar{d}d,~O_s= \bar{s}s \gap\textmd{or}\gap O_c= \bar{c}c,
\ee
depending on the quantity of interest. 
%For each flavour, the scalar matrix 
%element is parametrized by one form factor in the continuum, and we will 
%denote at zero momentum:
%\be
%\la N(p) | O_q (0) | N(p) \ra = g_{NNq} \bar{u}_N(p) u_N(p). 
%\ee
%%\la N(p') | O_q (0) | N(p) \ra = g_{NNq}(\mcQ^2) \bar{u}_N(p') u(p) 
%
Using the definitions of the two- and three-point functions of 
\eqs{eq:C2pts} and ~(\ref{eq:C3pts}), we will consider the ratio
\be\label{eq:ratio_def}
R_{O_q}(t_s,t_{\rm op}) = 
\frac{C^{\pm,O_q,\rm{vev}}_{N,\rm 3pts}(t_s,t_{\rm op})}{C^{+}_{N,\rm 2pts}(t, x_{\rm src}) } 
\ee
which is related to $y_N$ and will serve as our test quantity 
for the noise reduction technique to compute dis-connected 
contributions.
Our analysis has been performed on a $32^3\times 64$ volume with a lattice 
spacing of $a=0.0777(4)~\fm$ and a pion mass of approximately 
$380~\mev$. In order to improve the overlap between the ground 
state and the interpolating operators we use Gaussian smearing of the quark fields 
appearing in the interpolating fields. We use APE smearing of the gauge 
links involved in the Gaussian smearing, following the same 
strategy as in \cite{Dinter:2011jt,Drach:2010hy}.

\begin{figure}[htb]
\begin{center}
\hspace{-15pt}\includegraphics[width=0.8\textwidth,angle=\plotangle]{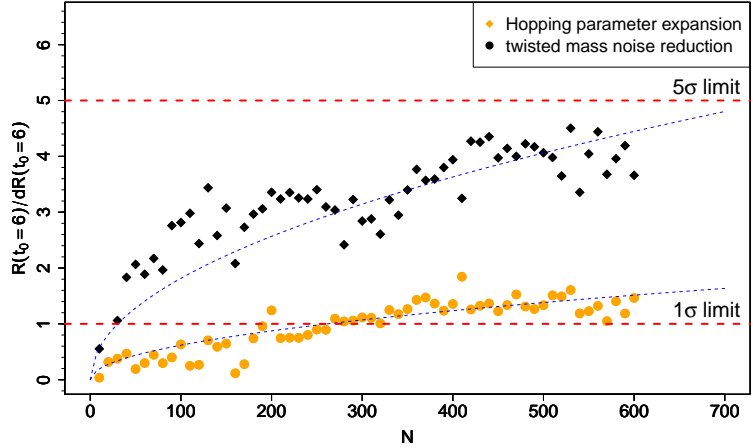}
\end{center}
\caption{Signal to noise ratio of the correlator ratio \eq{eq:ratio_def}, 
as a function of the number of configurations used, $N$,
for the variance noise reduction 
technique used in this paper compared to the more standard hopping parameter 
expansion technique. Note that here the source-sink separation 
is fixed to $t_s=12a$. Curves are only shown to guide the eyes. 
Our method allows to reach results at a $\sim 5\sigma$ significance 
level 
with a statistics that is $O(10)$ less than standard methods.} \label{fig:rel_err_scaling}
\end{figure}

We will use the powerful variance reduction method for twisted mass 
fermions introduced in ~\cite{Michael:2007vn,Boucaud:2008xu} and used to 
study the $\eta'$ meson in ~\cite{Jansen:2008wv}. 
In \fig{fig:rel_err_scaling}, we compare the efficiency of this technique 
with another, more standard noise reduction 
technique which relies on the hopping parameter expansion of the Dirac operator. 
This latter technique can be used for any discretization, and has been 
introduced in \cite{McNeile:2000xx}. We refer the interested reader to
appendix B. of ref.~\cite{Boucaud:2008xu} for the implementation 
of the method for twisted mass fermions. The 
here used new variance 
noise reduction technique improves the signal to noise ratio by a 
factor $\sim 3$ after having fixed all other parameters. 
In particular, as can be seen in fig.~\ref{fig:rel_err_scaling} with a statistics 
of $O(1000)$ gauge field configurations we can reach a result that 
is at the $5\sigma$ level different from zero, while the standard
method with such a statistics gives only a result at 
the $1\sigma$ level. 

Using our improved noise reduction technique for dis-connected contributions, 
we have 
first performed an investigation of the light quark 
content of the nucleon, i.e. $\sigma_{\pi N}$ of eq.~(\ref{eq:sigma_terms_FH}). 
Here, it turns out that the connected
part of the relevant 3-point function are by far dominating and that
the dis-connected contribution can be safely neglected. 
Although such a fact might have been expected, our dedicated 
investigation is certainly important to confirm this. 

\begin{figure}
%\vspace{-1cm}
\hspace{0.5cm}
\includegraphics[width=0.45\textwidth]{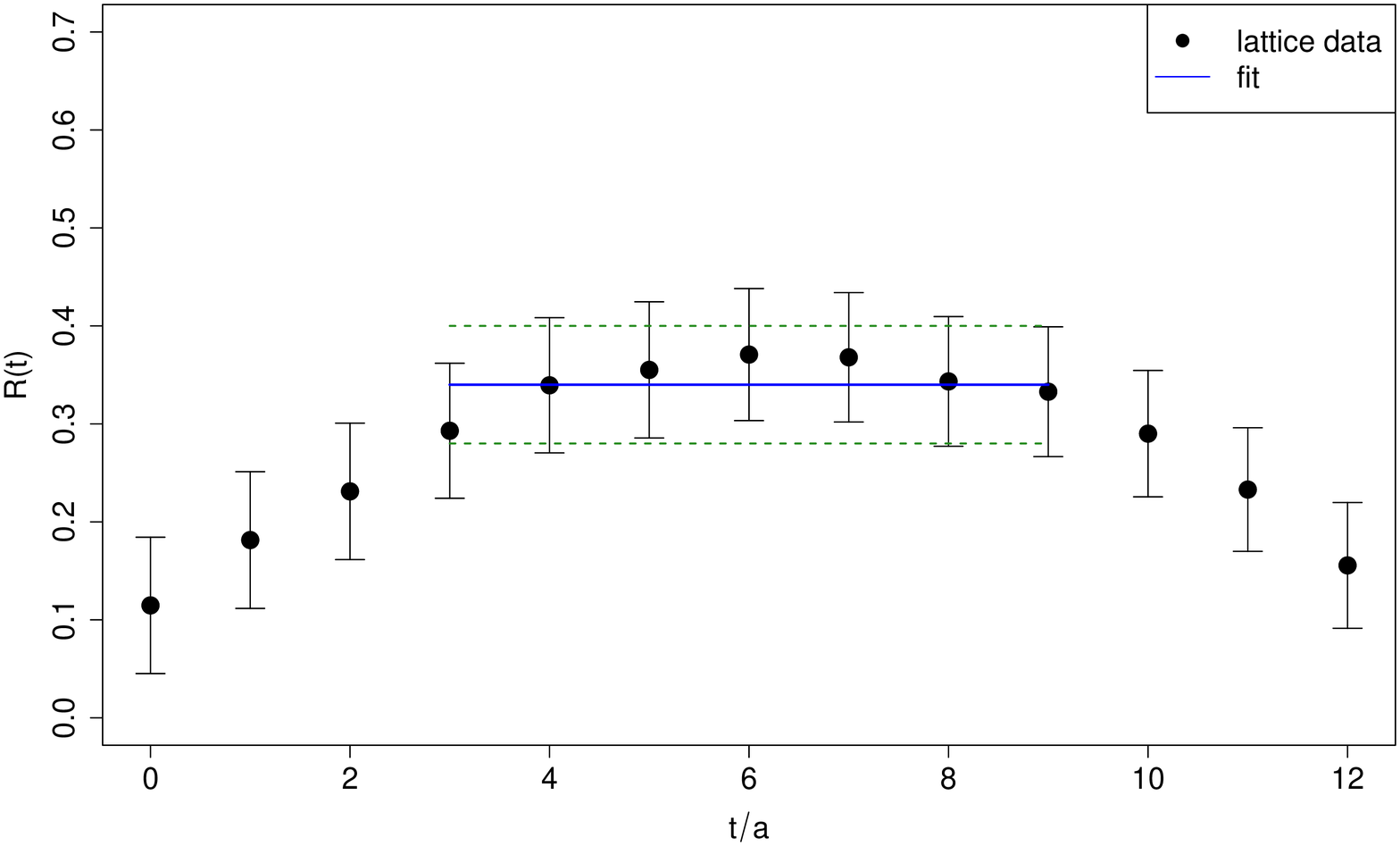}
\vspace{-1cm}
    \includegraphics[width=0.45\textwidth]{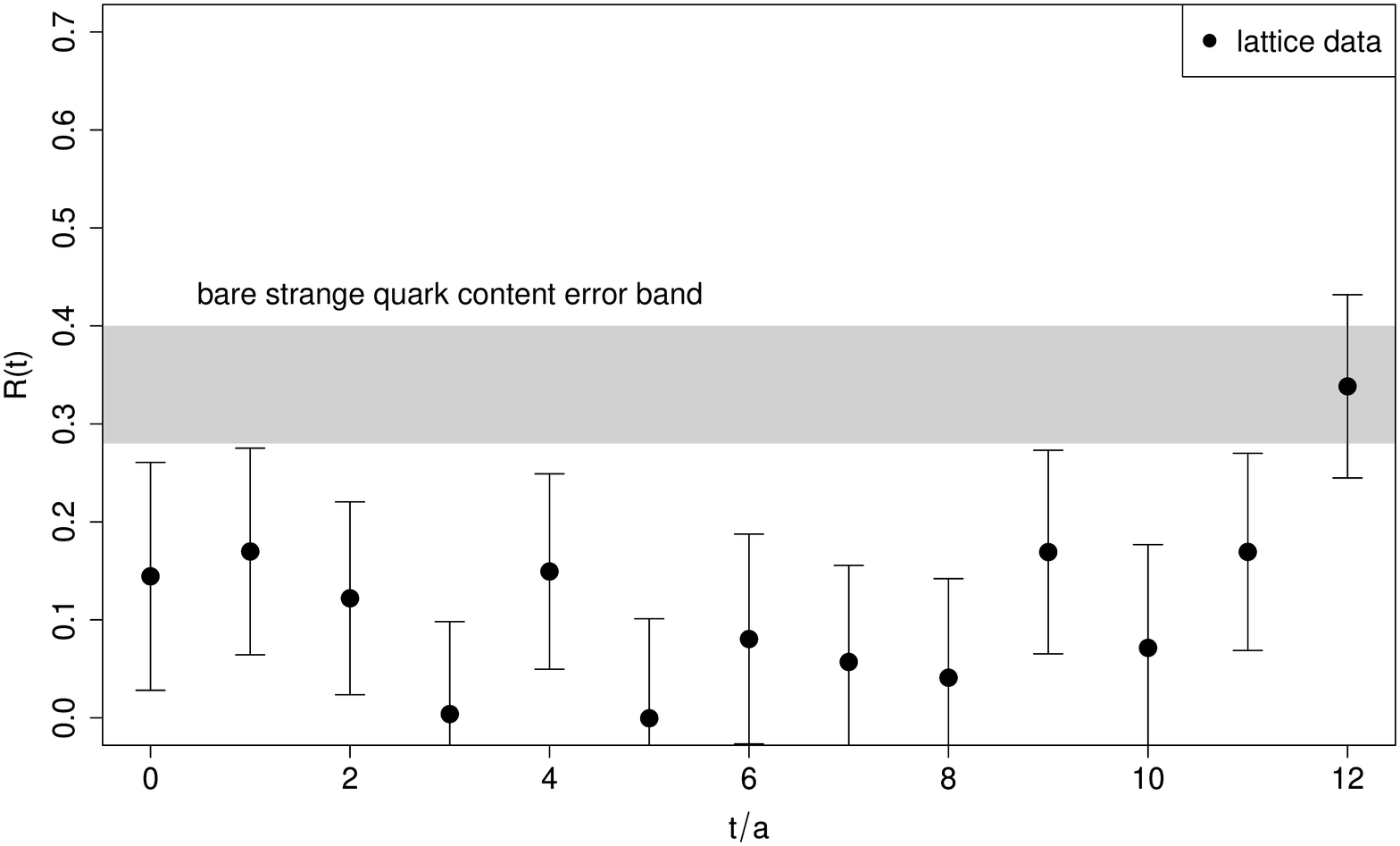}
\vspace{1cm}
\caption{In the left graph, 
we plot the bare ratio of \eq{eq:ratio_def} for the
strange quark. We used
$842$ configurations for this analysis. 
Note that the signal is 
$5\sigma$ away from $0$.
In the right graph, 
we show results for the charm content using the same statistics. While certainly 
no plateau can be identified, the charm quark content is clearly not larger than 
the strange quark content of the nucleon. As in fig.~\ref{fig:rel_err_scaling},  
$t_s=12a$. \label{fig:plateau_strange}}
\end{figure}

To estimate systematic effects coming from the excited states, the 
final number for $\sigma_{\pi N}$ 
is then obtained from a distribution of many 
fit results.  
From the (weighted) mean and variance we then obtain 
\be
\sigma_{\pi N}(m_{\rm{PS}} \approx 380 \mev ) = 150 (1)(10) \mev\; ,
\ee
where the first error is statistical and the second gives the systematic
uncertainties.
We then proceeded to compute the ratio $R$ of eq.~(\ref{eq:ratio_def})
for the strange and charm quark scalar operator. 
In \fig{fig:plateau_strange} (left), we show the plateau corresponding to the 
strange quark. 
The plateau value appears to be $5.7$ $\sigma$ away from zero showing
that we indeed can obtain a rather accurate 
value for the strange quark content of the nucleon.

We also carried out a first study 
of the charm quark content of the nucleon. Here it is most 
important that we can use gauge field configurations from 
$N_f=2+1+1$ simulations that take  
into account also the charm quark degree of freedom. We show in 
\fig{fig:plateau_strange} (right) our results for the charm quark.
Unfortunately, there is no plateau signal visible. 
The gray band in \fig{fig:plateau_strange} (right) represents our results 
for the strange quark. The graph then suggests that 
the charm 
quark content of the nucleon cannot be large compared to the 
strange quark as one might expect. 

\section{Conclusion}

\begin{figure}[htb]
\begin{center}
\hspace{-15pt}\includegraphics[width=0.8\textwidth,angle=\plotangle]{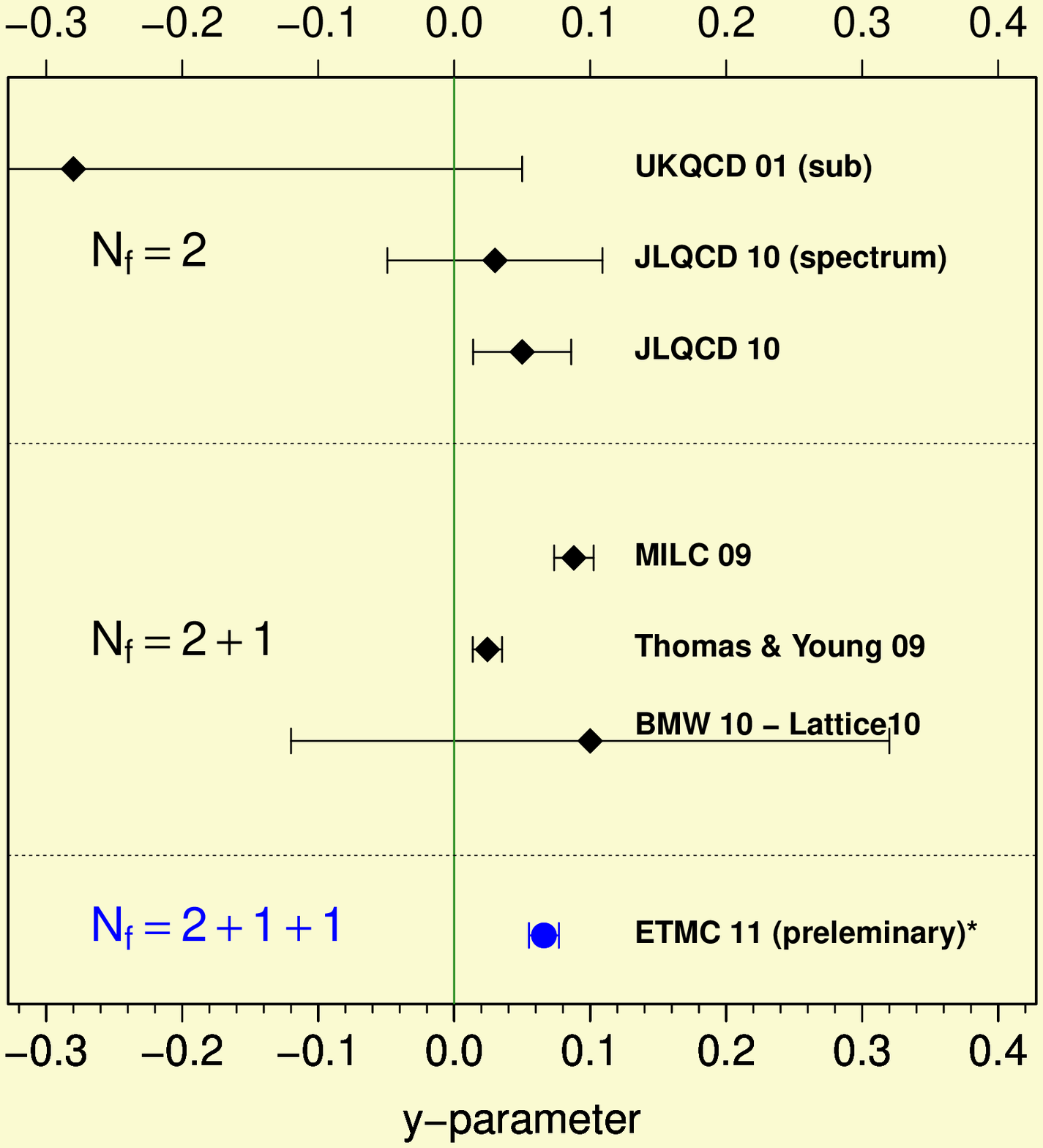}
\end{center}
\caption{Compilation of presently available results for the 
y-parameter $y_N$ of eq.~(\ref{eq:y_para}).}\label{fig:summary_y}
\end{figure}

In this proceedings contribution we have computed the scalar 
quark content of the nucleon for all four quarks of the first 
two generations. 
We find a value of the phenomenologically important quantity 
$y_N=0.066(11)(2)$, see eq.~(\ref{eq:y_para}).
This value is significantly smaller than predicted in chiral perturbation 
theory and indicates that the strange quark content is only 
a few percent. 
%In addition, we could obtain a value 
%of $y_N$ with an error of 17\%. 
In fig.~\ref{fig:summary_y} we give 
a summary of presently available values 
for $y_N$ demonstrating the precision of our result.   
We also could provide for the first time a bound for the charm 
quark content of the nucleon excluding that it is significantly larger 
than the strange quark content. 

As a present shortcoming, we mention that our calculations are  
performed at only one value of the lattice spacing, quark mass and volume. 
Thus we are lacking presently a good estimate of systematic effects
originating from varying these parameters of the simulations. 
Clearly, we will address these systematic errors in future computations which 
have already been started. 

\section*{Acknowledgments}
We thank all members of the ETM Collaboration for a very fruitful collaboration 
and many valuable discussions. In particular we want to thank R.~Frezzotti, G.~Herdoiza and 
G.~Rossi for essential discussions about the renormalization 
aspects. The HPC resources for this
 project have been made available by the computer centres of Barcelona, 
Groningen, J\"ulich, Lyon, Munich, Paris and Rome (apeNEXT), which we thank 
for enabling us to perform this work. This work has also been supported in part 
by the DFG Sonderforschungsbereich/Transregio SFB/TR9-03, and by 
GENCI (IDRIS - CINES), Grant 2010-052271.

\bibliographystyle{unsrt_nt}
\bibliography{paper}

\begin{thebibliography}{10}

\bibitem{Bertone:2004pz}
Gianfranco Bertone, Dan Hooper, and Joseph Silk.
\newblock {\em Phys.Rept.}, 405:279--390, 2005.

\bibitem{Ellis:2008hf}
John~R. Ellis, Keith~A. Olive, and Christopher Savage.
\newblock {\em Phys.Rev.}, D77:065026, 2008.

\bibitem{Giedt:2009mr}
Joel Giedt, Anthony~W. Thomas, and Ross~D. Young.
\newblock {\em Phys.Rev.Lett.}, 103:201802, 2009.

\bibitem{Young:2009ps}
Ross~D. Young and Anthony~W. Thomas.
\newblock {\em Nucl.Phys.}, A844:266C--271C, 2010.

\bibitem{Feynman:1939zz}
R.~P. Feynman.
\newblock {\em Phys. Rev.}, 56:340--343, 1939.

\bibitem{dinter}
S.~Dinter et~al.
\newblock in preparation.

\bibitem{Ohki:2009mt}
H.~Ohki, S.~Aoki, H.~Fukaya, S.~Hashimoto, T.~Kaneko, et~al.
\newblock {\em PoS}, LAT2009:124, 2009.

\bibitem{Ohki::2010cw}
H.~Ohki, S.~Aoki, H.~Fukaya, S.~Hashimoto, T.~Kaneko, et~al.
\newblock 2010.
%\newblock * Temporary entry *.

\bibitem{Collins:2010gr}
Sara Collins, Gunnar Bali, Andrea Nobile, Andreas Schafer, Yoshifumi Nakamura,
  et~al.
\newblock {\em PoS}, LATTICE2010:134, 2010.
%\newblock * Temporary entry *.

\bibitem{Bali:2009dz}
Gunnar Bali, Sara Collins, and Andreas Schafer.
\newblock {\em PoS}, LAT2009:149, 2009.

\bibitem{Freeman:2009pu}
Walter Freeman and Doug Toussaint.
\newblock {\em PoS}, LAT2009:137, 2009.

\bibitem{Durr:2010ni}
S.~Durr, Z.~Fodor, J.~Frison, T.~Hemmert, C.~Hoelbling, et~al.
\newblock {\em PoS}, LATTICE2010:102, 2010.
%\newblock * Temporary entry *.

\bibitem{Michael:2001bv}
Christopher Michael, C.~McNeile, and D.~Hepburn.
\newblock {\em Nucl.Phys.Proc.Suppl.}, 106:293--295, 2002.

\bibitem{Young:2009zb}
R.D. Young and A.W. Thomas.
\newblock {\em Phys.Rev.}, D81:014503, 2010.

\bibitem{Frezzotti:2000nk}
Roberto Frezzotti, Pietro~Antonio Grassi, Stefan Sint, and Peter Weisz.
\newblock {\em JHEP}, 0108:058, 2001.

\bibitem{Frezzotti:2003ni}
R.~Frezzotti and G.~C. Rossi.
\newblock {\em JHEP}, 08:007, 2004.

\bibitem{Frezzotti:2003xj}
R.~Frezzotti and G.~C. Rossi.
\newblock {\em Nucl. Phys. Proc. Suppl.}, 128:193--202, 2004.

\bibitem{Frezzotti:2004wz}
R.~Frezzotti and G.~C. Rossi.
\newblock {\em JHEP}, 10:070, 2004.

\bibitem{Alexandrou:2011nr}
C.~Alexandrou, J.~Carbonell, M.~Constantinou, P.A. Harraud, P.~Guichon, et~al.
\newblock {\em Phys.Rev.}, D83:114513, 2011.

\bibitem{Feng:2009ij}
Xu~Feng, Karl Jansen, and Dru~B. Renner.
\newblock {\em Phys.Lett.}, B684:268--274, 2010.

\bibitem{Feng:2010es}
Xu~Feng, Karl Jansen, and Dru~B. Renner.
\newblock {\em Phys.Rev.}, D83:094505, 2011.

\bibitem{Feng:2011zk}
Xu~Feng, Karl Jansen, Marcus Petschlies, and Dru~B. Renner.
\newblock 2011.

\bibitem{Baron:2009wt}
R.~Baron et~al.
\newblock {\em JHEP}, 1008:097, 2010.

\bibitem{Baron:2010bv}
R.~Baron et~al.
\newblock {\em JHEP}, 06:111, 2010.

\bibitem{Chiarappa:2006ae}
T.~Chiarappa et~al.
\newblock {\em Eur. Phys. J.}, C50:373--383, 2007.

\bibitem{Baron:2010th}
Remi Baron et~al.
\newblock {\em Comput.Phys.Commun.}, 182:299--316, 2011.

\bibitem{Dinter:2011jt}
Simon Dinter, Constantia Alexandrou, Martha Constantinou, Vincent Drach, Karl
  Jansen, et~al.
\newblock {\em PoS}, LATTICE2010:135, 2010.
%\newblock * Temporary entry *.

\bibitem{Drach:2010hy}
Vincent Drach, Karl Jansen, Jaume Carbonell, Mauro Papinutto, and Constantia
  Alexandrou.
\newblock {\em PoS}, LATTICE2010:101, 2010.
%\newblock * Temporary entry *.

\bibitem{Michael:2007vn}
Christopher Michael and Carsten Urbach.
\newblock {\em PoS}, LAT2007:122, 2007.

\bibitem{Boucaud:2008xu}
Ph. Boucaud et~al.
\newblock {\em Comput. Phys. Commun.}, 179:695--715, 2008.

\bibitem{Jansen:2008wv}
K.~Jansen, Christopher Michael, and C.~Urbach.
\newblock {\em Eur.Phys.J.}, C58:261--269, 2008.

\bibitem{McNeile:2000xx}
Craig McNeile and Christopher Michael.
\newblock {\em Phys.Rev.}, D63:114503, 2001.

\end{thebibliography}

\end{document}